\documentclass[aps,prl,twocolumn,superscriptaddress]{revtex4}
\usepackage{amsmath}
\usepackage{graphicx}
\usepackage{dcolumn}
\usepackage{bm}
\usepackage{amsfonts}
\makeatletter
\def\captionof#1#2{{\def\@captype{#1}#2}}
\makeatother

\newcommand{\rmd}{{\rm d}}
\newcommand{\rmi}{{\rm i}}
\newcommand{\ud}[1]{\hspace{-0em}\mathrm{d}{#1}\;}

\begin{document}

\title{Analytic theory of Hund's metals: A renormalization group perspective}

\author{Camille Aron}
\affiliation{Department of Physics and Astronomy, Rutgers University, Piscataway, NJ 08854, USA}
\affiliation{Department of Electrical Engineering, Princeton University, Princeton, NJ 08544, USA}
\author{Gabriel Kotliar}
\affiliation{Department of Physics and Astronomy, Rutgers University, Piscataway, NJ 08854, USA}

\begin{abstract}
We study the emergence of quasiparticles in Hund's metals with an SU($M$)$\times$ SU($N$)-symmetric Kondo impurity model carrying both spin and orbital degrees of freedom. We show that the coupling of the impurity spin to the conduction electrons can be \emph{ferromagnetic}, notably for hole-doped iron pnictides.
We derive the weak-coupling renormalization group (RG) equations for arbitrary representations of SU($M$)$\times$SU($N$). A ferromagnetic spin coupling results in a protracted RG flow, accounting for the surprising particle-hole asymmetry that is observed in the iron-pnictide systems. We establish the low coherence scale $T_{\rm K}$ which depends on the filling through the impurity representation. We also discuss the temperature dependence of the spin and orbital susceptibilities. Finally, we argue that this mechanism explains the strong valence dependence of the coherence scale observed in dilute transition-metal magnetic alloys.
\end{abstract}

\maketitle

%\paragraph{Introduction to Hund's Metals.}
There is a renewed interest in a class of materials where strong electronic correlations, manifest in large mass renormalizations, arise from Hund's coupling rather than from the Hubbard $U$ term. Noticeable examples are the recently discovered iron-pnictides and chalcogenides high-temperature superconductors~\cite{HK,YHK}, ruthenates~\cite{Werner,Mravlje}, or other $4d$ transition-metal oxides~\citep{Medici}. 

A local approach seems a promising route for the understanding of Hund's metals. GW calculations support the idea that the self-energy at low energies has a purely local character~\cite{Jan}. LDA+DMFT studies, mapping the many-body problem to an impurity problem in a self-consistent determined environment, has provided a successful description of several materials in this class~\cite{Medici}. 
Hund's metals form a Fermi liquid below a coherence temperature which is remarkably low~\cite{HK}. The physical degrees of freedom at higher energies are fluctuating moments~\citep{Hansmann} which are observed in XES measurements~\cite{Gretarsson} and incoherent electronic excitations which are observed in their optical properties~\cite{Schafgans,Katsufuji,Lee}. 
 
Since the DMFT bath of the Hund's metals is relatively structureless at low energies, it is natural
to investigate this problem with a representative impurity model,
an SU($M$)$\times$SU($N$) generalization of the model introduced in Ref.~\cite{KotliarHauleYin}
by means of an analytical renormalization group analysis. 
The goal is to get analytical insights into 
 why is the coherence scale of Hund's metals
so low, and what are the physical parameters that control its value. 
This mystery dates back to the fifties 
when early investigations of the Kondo temperature $T_{\rm K}$ of dilute transition-metal magnetic alloys
revealed that $T_{\rm K}$ decreases dramatically as the $d$-shell filling approaches half-filling~\cite{Daybell,Schrieffer}. The renormalization group flows describe an interesting interplay of spin and orbital degrees of freedom, give new insights into why the spin and orbital susceptibility are so different and account for the surprising particle-hole asymmetry observed in the iron-pnictide systems. 

\paragraph{Model.} We study the impurity model described by the Hamiltonian $H_{\rm K} = H_{\rm bath} + H_{\rm int}$ where $H_{\rm bath} = \sum_{k,m,\sigma} \epsilon_k \, \psi^\dagger_{km\sigma} \psi_{km\sigma}$ describes the non-interacting conduction electrons $\psi_{k\sigma a}$ with momentum $k$. $\sigma = 1 \ldots N$ labels the spin of the electron and $m= 1 \ldots M$ labels its orbital. $M$ is the number of active orbitals in the shell (\textit{i.e.} $ M=3$ for $t_{2g}$ or $M=5$ for a full shell of $d$ electrons). The physical case for the spin sector is $N=2$ but we keep its value general.
We consider a dispersion $\epsilon_k$ corresponding to a flat density of states $\rho$ (we later set $\rho =1$ to simplify expressions) with large bandwidth $2D_0$.
The spin and orbital degrees of freedom of the impurity, $\boldsymbol{S}$ and $\boldsymbol{T}$, live respectively in faithful representations of SU($N$) and SU($M$) to be precised below.
The coupling of the impurity to the conduction electrons reads (summing over repeated indices)
\begin{align}
 H_{\rm int} &= 
 J_{\rm p}\, \psi^\dagger_{a\sigma}\psi_{a\sigma} + 
 J_0 \, S^\alpha ( \psi^\dagger_{m \sigma} \frac{{\sigma}^\alpha_{\sigma\sigma'}}{2} \psi_{m \sigma'} ) \label{eq:KondoModel}
 \\ 
 &\! \!\!\!\!\!\!\!\!\!\! + \!
 K_0 T^a ( \psi^\dagger_{m\sigma} \frac{{\tau}^a_{mm'}}{2} \psi_{m'\!\sigma} \! ) 
\! + \! I_0 S^\alpha T^a ( \psi^\dagger_{m\sigma} \frac{{\sigma}^\alpha_{\sigma\sigma'}}{2} \! \frac{{\tau}^a_{mm'}}{2} \psi_{m'\!\sigma'} \!), \nonumber 
%\\
%H_{\rm bath} &\equiv 
% \sum_{k,m,\sigma} 
% \epsilon_k \, \psi^\dagger_{km\sigma} \psi_{km\sigma}\;. 
\end{align}
with the local conduction electron $\psi_{m\sigma} \equiv \sum_k \psi_{km\sigma}$. $J_{\rm p}$, $J_0$, $K_0$, and $I_0$ are respectively the bare potential, spin-spin, orbital-orbital, and spin-orbital Kondo coupling constants. 
$\sigma^\alpha$ ($\alpha=1\ldots N^2-1$) and $\tau^a$ ($a=1\ldots M^2-1$) are the generators of SU($N$) and SU($M$) respectively in their fundamental representations.
They obey the Lie algebra commutation relations and are normalized such that $\mbox{Tr}\left[\sigma^\alpha \sigma^\beta \right] = 2 \delta_{\alpha \beta}$ and $\mbox{Tr}\left[\tau^a \tau^b \right] = 2 \delta_{ab}$.
For SU(2) and SU(3), they correspond to the Pauli and Gell-Mann matrices respectively.

\begin{figure}[!t]
\centerline{
\includegraphics[width=0.60\columnwidth]{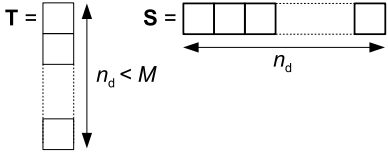}
\vspace{-0.2cm}
}
\caption{\label{fig:young}\footnotesize Young's tableaux of the representations of $\boldsymbol{T}$ and $\boldsymbol{S}$ in our class of Hund's metals. A single box represents a fundamental spin of SU($M$) or SU($N$).}
\end{figure}

We consider the case of Hund's metals with valences $n_d$ less than half-shell filling. Above half-shell capacity, one can perform a particle-hole transformation, before 
generalizing from SU(2) to SU($N$). We denote the distance from half-filling by $d\equiv M-n_d \geq 1$. 
The effect of a strong Hund's coupling is to maximize the impurity spin, therefore we take $\boldsymbol{S}$ as the generators of the totally symmetric representation of $n_d$ fundamental SU($N$) spins and $\boldsymbol{T}$ to live in the totally antisymmetric representation composed of $n_d <M$ fundamental SU($M$) isospins.
These representations correspond to the Young's tableaux in Fig.~\ref{fig:young}.
Notice that at exactly $1/N$-filling, \textit{i.e.} $n_d=M$, the orbital isospin is a singlet state (scalar representation) and the model reduces to an $M$-channel Coqblin-Schrieffer model with a totally antisymmetric spin representation~\cite{ParcolletAntisym}.

The Kondo model in Eq.~(\ref{eq:KondoModel}) can be derived, \textit{via} a canonical Schrieffer-Wolff transformation, from the large interaction limit of the SU($M$)$\times$SU($N$)-symmetric Anderson impurity Hamiltonian~\cite{Parmenter}, $H_{\rm AIM} = H_{\rm imp} + H_{\rm hyb} + H_{\rm bath}$ with
\begin{align} 
H_{\rm imp} &\equiv \epsilon_d \, n_d 
+ \frac{1}{2} \,
 \!\!\! \sum_{mnpq,\, \sigma\sigma'} \!\!\!\! 
 U_{mnpq} \,
 d_{m\sigma}^\dagger d_{n\sigma'}^\dagger d_{p\sigma'} d_{q\sigma}\;, \\
H_{\rm hyb} &\equiv 
V
\sum_{k,m,\sigma} 
 \psi^\dagger_{km\sigma} d_{m\sigma} + {\rm H.c.}\;.
% \\
%H_{\rm bath} &\equiv 
% \sum_{k,m,\sigma} 
% \epsilon_k \, \psi^\dagger_{km\sigma} \psi_{km\sigma}\;. 
\end{align}
$d_{m\sigma}$ represents an impurity electron with spin $\sigma$ in the orbital $m$, $\epsilon_d$ is the energy level and $n_d \equiv \sum_{m\sigma} d_{m\sigma}^\dagger d_{m\sigma}$. The second term of $H_{\rm imp}$ encodes both Coulombic repulsion and Hund's coupling with $U_{mnpq} \equiv U \delta_{mq}\delta_{np} + J_{\rm H} \delta_{mp} \delta_{nq}$. $H_{\rm hyb}$ is the hybridization with the conduction electrons.

In the large interaction limit, $U \gg D_0 \gg J_{\rm H} \gg V$, the charge degrees of freedom of the Anderson impurity are frozen, and the nominal valence of the impurity is identified to $n_d$. The states of the impurity carry an SU($N$) spin $\boldsymbol{S}$ and an orbital SU($M$) isospin $\boldsymbol{T}$ interacting according to $H_{\rm int}$, with the Kondo couplings~\cite{Supplementary}
%obtained by performing from $H_{\rm AIM}$ to $H_{\rm K} = H_{\rm int} + H_{\rm bath}$
\begin{align}
 J_{\rm p} &= \frac{1}{MN} \left[ \frac{n_d}{\Delta E^-} - \frac{M-n_d}{n_d + 1} \frac{N+n_d}{\Delta E^+} \right] V^2\;, \\
 J_0 &= \frac{2}{M} \left[ \frac{1}{\Delta E^-} - \frac{M-n_d}{n_d + 1} \frac{1}{\Delta E^+} \right] V^2\;, \label{eq:J0} \\
 K_0 &= \frac{2}{N} \left[ \frac{1}{\Delta E^-} + \frac{N+n_d}{n_d+1} \frac{1}{\Delta E^+} \right] V^2\;, \\
 I_0 &= 4 \left[ \frac{1}{n_d} \frac{1}{\Delta E^-}+ \frac{1}{n_d + 1} \frac{1}{\Delta E^+}\right] V^2\;, \label{eq:I0}
\end{align}
in which the virtual charge excitation energies to the $n_d\pm 1$ valence states, $\Delta E^+ \approx \epsilon_d + n_d U$ and $\Delta E^- \simeq - \epsilon_d - (n_d-1)U$, are both positive if $\epsilon_d = -(n_d-1 + \alpha) U$ with $\alpha \in\, ]0,1[$. 
The minus sign in front of the second term of $J_0$ above implies that, depending on the value of $\epsilon_d$, $J_0$ can be significantly smaller than the other couplings, and even ferromagnetic, $J_0 < 0$.
For $\alpha > \alpha^* \equiv (n_d+1)/(M+1)$ virtual transitions to valence $n_d+1$ dominate and $J_0$ is ferromagnetic. 
For iron pnictites or ruthenates which have $M=5$ or $M=3$ with valences one unit larger than half-filling, a preliminary particle-hole transformation yields $n_d=4$ or $n_d=2$, respectively, and thus hole doping favors a ferromagnetic $J_0$.

The possibility of such a ferromagnetic spin coupling, is a consequence of the large Hund's coupling encoded in our choice of representations. Indeed, setting $J_{\rm H} = 0$ yields positive Kondo couplings with $J_0 = 2/M \left[ 1/\Delta E^+ + 1/\Delta E^- \right] V^2$ and $n_d I_0 = 2 M J_0 = 2 N K_0$~\cite{Supplementary}. In this case, the model reduces to a single-channel Coqblin-Schrieffer model, with a single Kondo coupling $\mathcal{J}_0$ between the conduction electrons and an impurity spin living in the totally antisymmetric representation of SU($M\times N$) and composed of $n_d$ electrons.

\begin{figure}[!t]
\centerline{
\includegraphics[width=0.65\columnwidth]{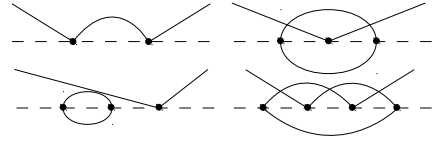}
\vspace{-0.2cm}
}
\caption{\label{fig:diag}\footnotesize Above: second and third-order non-parquet diagrams contributing to the RG equations (\ref{eq:beta_J})-(\ref{eq:beta_I}). Below: third-order renormalization of the wave function and a fourth order diagram.}
\end{figure}
\paragraph{RG equations.}
To study the physical properties of the Kondo model, we use a poor man's scaling approach at zero temperature~\cite{Anderson,Solyom}. This consists in reducing the bandwidth by perturbatively integrating over the degrees of freedom of those conduction electrons with an energy in the edge $\delta D$ of the band and requiring that the physics remains invariant.
The corresponding renormalization of the couplings is given by the so-called $\beta$ functions, $\beta_i \equiv \rmd J_i / \rmd\ln D$ with $J_i=J,K,I$, together with the initial conditions set by the bare couplings, $J(D_0) = J_0$, $K(D_0) = K_0$ and $I(D_0) = I_0$. The expansion of the $\beta$ functions to any order in the couplings can be expressed in terms of $C_n^S$ and $C_n^T$, the eigenvalues of $n$-th order Casimir invariants of the representations of $\boldsymbol{S}$ and $\boldsymbol{T}$ respectively
\footnote{$C_n^S \, \mathbb{I}_S \equiv \mbox{Tr}[ \frac{\sigma^{\{ \alpha_1}}{2} \ldots \frac{\sigma^{\alpha_n \}}}{2}] S^{\alpha_1} \ldots S^{\alpha_n} $ where $\mathbb{I}_S$ is the identity in the representation of $\boldsymbol{S}$ and
$\{a_1 \ldots a_n \}$ stands for the sum over all permutations weighted by $1/n!\,$.}. 
Up to third order, we obtain (see Fig.~\ref{fig:diag})
\begin{align}
%\frac{{\rm d} J_{\rm p}}{{\rm d}\ln D } &= 0 \\ 
\beta_J &= -\displaystyle \frac{N}{2} \left( 1-\frac{M}{2} J \right) \left( J^2 + \frac{C_2^T}{2M} I^2 \right)
+ \ldots
\;, \label{eq:beta_J}
\\
\beta_K &= -\frac{M}{2} \left( 1-\frac{N}{2} K \right) \left( K^2 + \frac{C_2^S}{2N} I^2 \right) + \ldots \;, \label{eq:beta_K}\\
 \beta_I &= -\frac{MN}{4} \left[
 \left( \frac{4}{M}J + \frac{4}{N}K - J^2 - K^2 \right) I 
 \label{eq:beta_I} \right. \\
 & + \!\!\left. \left(\frac{ C_3^T}{M C_2^T}\! + \!\frac{ C_3^S}{N C_2^S} \right)\! I^2 
 \!+ \!\left( \frac{1}{4}\! -\!\frac{C_2^T}{2M}\! -\! \frac{C_2^S}{2N} \right)\! I^3 \right] + \ldots\;.
 \nonumber
\end{align}
For the sake of generality, we gave the $\beta$ functions for $\boldsymbol{S}$ and $\boldsymbol{T}$ living in \emph{arbitrary} faithful representations of SU($N$) and SU($M$). These equations have a broad range of applicability since the spin-orbital Kondo effect can be realized in different settings such as bilayer graphene~\cite{Maxim} or nanoscale devices~\cite{carbon}. 
We shall later return to our particular model by specifying the Casimirs for the Hund's metals.
  
We discarded the flow of potential scattering since it does not renormalize the other couplings.
We also discarded the flow of quadrupolar spin-orbital interactions generated by the pertubative expansion but  not initially present in $H_{\rm int}$.
For example, the term in 
$I^2 \left(\boldsymbol{S} \cdot \boldsymbol{\sigma} \right) \left( \boldsymbol{Q}\cdot \boldsymbol{\tau} \right)$
 with $Q^c \equiv \{T^a,T^b \} \mbox{Tr}\left[ \tau^{\{a} \tau^b \tau^{c\}} \right]$ was projected on $\left( \boldsymbol{S} \cdot \boldsymbol{\sigma} \right) \left( \boldsymbol{T} \cdot \boldsymbol{\tau} \right)$~\cite{Supplementary}~\footnote{In the fundamental representations, quadrupolar terms are simply absent since one necessarily has $\boldsymbol{Q} \propto\boldsymbol{S}$ owing to the fact that $\{\mathbb{I}, S^a ; a=1\ldots N^2-1 \}$ is a complete basis of the SU($N$) operators.}.
Notice that up to the third order in the couplings, the representation-dependent factors in the $\beta$ equations vanish for $I=0$, as expected for two uncoupled multi-channel Coqblin-Schrieffer models.

The six fixed points of the RG Eqs.~(\ref{eq:beta_J})-(\ref{eq:beta_I}) are easily identified as
$(i)~J=K=I=0$, the non-interacting fixed point,
$(ii)~J=J^*\equiv 2/M, K=I=0$, the intermediate-coupling fixed point of the $N$-channel SU($M$) Coqblin-Schrieffer model, and
$(iii)~K=K^*\equiv 2/N, J=I=0$, the one of the $M$-channel SU($N$) Coqblin-Schrieffer model.
$(i)$, $(ii)$ and $(iii)$ are unstable against $J_0>0$ or $K_0>0$ and, as long as $I_0=0$, the RG flows to the fixed point $(iv)~J=J^*, K=K^*, I=0$ which corresponds to the fixed point of two uncoupled multi-channel Coqblin-Schrieffer models and the low-energy physics is dominated by the one with the smallest Kondo scale.
As soon as $I_0\neq 0$, the fixed points $(i)$-$(iv)$ are all unstable and the RG eventually flows towards $(v) ~J=J^*, K=K^*, I=I^*_-$ or $(vi)~J=J^*, K=K^*, I=I^*_+$ depending on the sign of $I_0$. 
Here, 
%\begin{align} \label{eq:I*}
$ I_\pm^* \equiv 
\left( a \pm \sqrt{bc+a^2} \right)/b,$
%\end{align}	
with $a\equiv {C_3^S}/{N C_2^S} + {C_3^T}/{M C_2^T}$, $b\equiv {C_2^S}/{N} + {C_2^T}/{M} -1/2$ and $c\equiv 8\left( {1}/{N^{2}} + {1}/{M^{2}} \right)$.
Contrary to $(i)$-$(iv)$, the locations of the fixed points $(v)$ and $(vi)$ and the RG flows around them, depend on the representations of the impurity spin $\boldsymbol{S}$ and isospin $\boldsymbol{T}$.

%\paragraph{Validity.}
The pertubative expansion of the $\beta$ functions are only reliable around the non-interacting fixed point $(i)$ and one must be careful before assigning a physical meaning to $(v)$ and $(vi)$.
When both sectors, $\boldsymbol{S}$ and $\boldsymbol{T}$, are in their fundamental representation, $C^S_2={(N^2-1)}/{2N}$ and $C^S_3 = (N^2-1)(N^2-4)/4N^2$ (and similar expressions for $C_2^T$ and $C_3^T$), one recovers the $\beta$ equations derived in Ref.~\cite{Kuramoto}.
For SU(2)$\times$SU(2), $(v)$ and $(vi)$ with $I^*_\pm = \pm 4$ are known to be artefacts of the third-order expansion, and the correct fixed point is a strong-coupling fixed point at $I$, $J$, $K\to\infty$.
For arbitrary $M$ and $N$,  $(v)$ with $I_-^* = -4 \frac{N^2 + M^2}{N^2M^2-N^2 -M^2}$ is well defined at large $N$ and $M$ and it was conjectured to be stable for all $N$ and $M$ except for $N=M=2$~\cite{Kuramoto}. On the other hand, $(vi)$ with $I_+^*=4$ lies out of the scope of the pertubative analysis. Kuramoto argued that, similarly to the SU(2)$\times$SU(2) case, it should be replaced by a strong-coupling fixed point. This is particularly clear at the special values of couplings $2MJ= 2NK = I$ for which the model reduces to the SU($M\times N$) Coqblin-Schrieffer model which has only a non-interacting and a strong-coupling fixed point.

\paragraph{RG flow of Hund's metals.}
We now return to Hund's metals by working with the totally symmetric and antisymmetric representations introduced before (see Fig.~\ref{fig:young}).
The Casimirs read
$C_2^S = (N-1)n_d (N+n_d)/2N$, $C_3^S = (N-2)(N-1)n_d(N+n_d)(N+2n_d)/4N^2$, 
$C_2^T = (M+1)n_d (M-n_d)/2M$, and $C_3^T = (M+2)(M+1)n_d (M-n_d)(M-2n_d)/4M^2$~\cite{Okubo}.
Henceforth, we work in the large-$N$ large-$M$ limit while keeping both the ratio $q \equiv M/N$ and the distance to $1/N$-filling, $d\equiv M-n_d \geq 1$, finite.
In this limit, the fixed points $(v)$ and $(vi)$ are located at
\begin{align}
I_-^* \simeq -\frac{4}{NM}\;, \mbox{ and } I_+^* \simeq \frac{4}{M}.
\end{align}
both lying out of the convergence domain of the perturbative expansion~\footnote{A rapid inspection shows that the higher order terms in the expansion of $\beta_I$ scale as $N^{2n-2} I^n$ for $n\geq 4$ (see \textit{e.g.} the fourth order term in Fig.~\ref{fig:diag}). This indicates that the pertubative expansion converges if $|I|\ll1/N^2$.}.
Based on numerical renormalization group results~\cite{Hewson_asym} and numerical findings~\cite{KotliarHauleYin}, we conjecture that the flow towards $(vi)$ at ($J^*,K^*, I_+^*$) should be understood as a flow to strong coupling and we use $(vi)$ only to estimate the energy scale at which the Fermi-liquid coherence is restored.

The RG Eqs.~(\ref{eq:beta_J})-(\ref{eq:beta_I}) can be solved numerically with arbitrary bare couplings $J_0$, $K_0$ and $I_0$ as initial conditions. Below, we illustrate how the RG trajectories depend on $J_0$ by solving them analytically in three regimes: weakly ferromagnetic $|J_0| \lesssim K_0$, strongly ferromagnetic $|J_0| \gg K_0$ and strongly antiferromagnetic $J_0 \gg K_0$. % For each scenario, we compute the low-energy coherence scale and we discuss the spin and orbital susceptibilities.
Not all these regimes of couplings can be reached from the strong-coupling limit of the multi-band Anderson model, see Eqs.~(\ref{eq:J0})-(\ref{eq:I0}), so that the Kondo model is a more general low-energy model. This is justified because in actual materials there are additional ligand bands contributing to the Kondo couplings.

In the large-$M$ large-$N$ limit and to quadratic order, the RG equations read
\begin{align}
\beta_J &= - {N}/{2} \left[ J^2 + d/4 \ I^2 \right]
+ \ldots
\;, \label{eq:beta_J2}
\\
\beta_K &= - {M}/{2} \left[ K^2 + N q(1+q)/4 \ I^2 \right] + \ldots \;, \label{eq:beta_K2}\\
 \beta_I &= -NI [J+qK+ q^2 N /4\ I] + \ldots\;.
 \label{eq:beta_I2}
\end{align}
To discuss different types of RG flow, we introduce
%\begin{align} \nonumber
$ T_{\rm K}^K \approx  \exp(-2/M \! K_0) D_0$, $T_{\rm K}^I \approx \exp(-4/M^2\! I_0) D_0$
and $T_{\rm K}^J \approx \exp(-2/N \! J_0) D_0$ if $J_0 >0$
%\end{align}
which are the intrinsic Kondo scales in absence of cross-terms in Eqs.~(\ref{eq:beta_J2})-(\ref{eq:beta_I2}).
Below, we consider the spin-orbital coupling as the smallest coupling by assuming the hierarchy $T_{\rm K}^I < T_{\rm K}^K$.

\begin{figure}[!t]
\centerline{
\includegraphics[width=0.95\columnwidth]{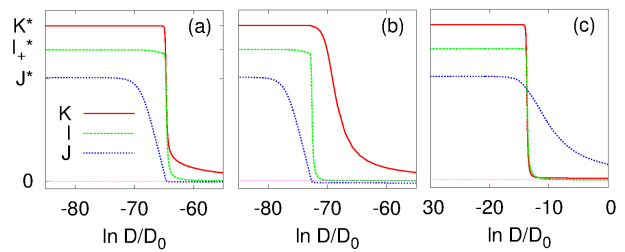}
\vspace{-0.3cm}
}
\caption{\label{fig:flow}\footnotesize  (Color online)
Numerical RG flow starting from
(a) weakly ferromagnetic $|J_0| = 10^{-3} \lesssim K_0$, (b) strongly ferromagnetic $|J_0| = 10^{-1} \gg K_0$, and (c) strongly antiferromagnetic $J_0 = 10^{-2} \gg K_0$. ($q=3/2$, $d=1$, $K_0 = 10^{-3}$, $I_0 = 10^{-5}$, $N=20$).}
\end{figure}

Let us first examine the case of a weakly ferromagnetic spin coupling, $J_0 < 0$, with $|J_0| \lesssim K_0$.
See Fig.~\ref{fig:flow}(a).
At high energies, the terms involving $I$ in the RG Eqs.~(\ref{eq:beta_J2}) and (\ref{eq:beta_K2}) can be neglected, thus spin and orbital degrees of freedom are decoupled.
The antiferromagnetic coupling $K$ of the totally antisymmetric SU($M$) pseudo-spin approaches the non-Fermi-liquid fixed point $(ii)$ controlled by the Kondo scale $T_{\rm K}^K$ and the scaling exponent $\Delta_K \equiv \rmd\beta_K/\rmd K \approx q$~\cite{ParcolletAntisym} while the ferromagnetic coupling $J$ of the totally symmetric SU($N$) spin slowly flows to weak coupling with an exponent $\Delta_J \equiv \rmd\beta_J/\rmd J \approx 0$.
At energy scales of the order of $T_{\rm K}^K$, $J$ is still ferromagnetic while $K$ reaches its fixed point, $K(T_{\rm K}^K) \approx K^*$. Below $T_{\rm K}^K$, $K^*$ controls $\beta_I \approx - M K^* I < 0$ and the spin-orbital coupling renormalizes to strong coupling, $I(T_{\rm K}^K) \approx I_+^*$.
Then, the $I^2$ term in Eq.~(\ref{eq:beta_J2}) drives the growth of $J$ which crosses over from a ferromagnetic to an antiferromagnetic value.
The integration of Eq.~(\ref{eq:beta_J2}) provides an estimate of the typical energy scale $T_{\rm K}$ at which $J \to J^*$, \textit{i.e.} at which the strong-coupling regime establishes,
\begin{align} \label{eq:KondoT}
 T_{\rm K}(d) \approx \exp\left(-{q}/{d} \right) T^K_{\rm K}\;.
\end{align}
Note that Eq.~(\ref{eq:KondoT}) is still valid for a relatively small antiferromagnetic coupling, as long as $T_{\rm K}^J < T_{\rm K}^K$ or $T_{\rm K}^J < T_{\rm K}^I$.
In agreement with the experimental and numerical evidence, $T_{\rm K}$ is found to decrease as one approaches $1/N$-filling (\textit{i.e.} as $d$ gets smaller).
At a more formal level, $T_{\rm K}$ depends \emph{explicitly} on the representations of the spin and the orbital isospin. This is unlike the typical Kondo scales emerging in Kondo models without spin-orbital coupling which can only depend \emph{implicitly} on the representations through the bare couplings ($\textit{e.g.}$ $T_{\rm K}^K$ or $T_{\rm K}^J$ given above).

Let us now discuss the scenario with large ferromagnetic coupling $|J_0| \gg K_0$.
See Fig.~\ref{fig:flow}(b).
As seen in Eq.~(\ref{eq:beta_I2}), $J$ controls the renormalization of $I$ as long as $K \ll |J|$ and $\beta_I \approx - N J I > 0$. Thus, $I$ first slowly renormalizes to weak coupling and reaches values on the order of $I_0' \equiv q^2 I_0 K_0^2/|J_0|^2 \ll I_0$ at $T_{\rm K}^K$ (when $K\to K^*$).
The subsequent growth of $I$ to $I^*_+$ is therefore delayed by such a small initial value 
and $I$ escapes weak coupling at a scale $ I_0'^{-1/\Delta_I} T_{\rm K}^K < T_{\rm K}^K$ with $\Delta_I \equiv \rmd \beta_I/\rmd I \approx -2q$. In turn, this also delays the subsequent renormalization of $J$ to $J^*$.
The relevance of multi-channel Kondo physics for the intermediate asymptotics was conjectured and the operator responsible for the crossover to the Fermi liquid at low energies was identified in Ref.~\cite{Tsvelik}.

It is useful to contrast the scenarios above with the case of large antiferromagnetic coupling $J_0 \gg K_0$. See Fig.~\ref{fig:flow}(c). There, the RG flow is radically different as all three couplings reach strong coupling concomitantly at the scale set by $T_{\rm K}^J > T_{\rm K}^K$.

Finally, our results can also be compared to the case of the absence of Hund's coupling, $J_{\rm H} = 0$, for which the model reduces to the antiferromagnetic SU($M\times N$) Coqblin-Schrieffer model described before. There, all antiferromagnetic Kondo couplings are locked together by symmetry considerations and strong coupling is reached at energies on the order of $\exp\left[ -2/(MN\mathcal{J}_0) \right] D_0$.

\paragraph{Susceptibilities.}
The RG flow can be used to study physical observables such as the impurity spin and orbital static susceptibilities, $\chi_S$ and $\chi_T$ respectively. The temperature scaling equations derived in Ref.~\cite{Supplementary} have solutions 
\begin{align}
\chi_{S/T}(T) \sim \frac{1}{T} \exp\left( - \int_{T}^{D_0} \! \! \frac{\rmd D}{D} \, \gamma_{S/T}\left(J_i(D)\right) \right)\;, \label{eq:chiST}
\end{align}
with the functions $\gamma_S = MN (J^2 + C^T_2 I^2 / 2) / 2$ and $\gamma_T = MN (K^2 + C^S_2 I^2 / 2) / 2$.
Let us focus on the ferromagnetic case, $J_0 < 0$. At high temperatures $T \sim D_0$, the exponent above can be neglected and both susceptibilities follow a Curie law, \textit{i.e.} $1/T$. At temperatures down to $T_{\rm K}^K$, Eq.~(\ref{eq:chiST}) and the RG flow discussed above imply that the magnitude of $\chi_T$ is significantly smaller than the one of $\chi_S$. At $T_{\rm K}^K$, the orbital susceptibility crosses over to a strong-coupling regime and $\chi_T\to 0$ when $T\to 0$. 
In the weakly ferromagnetic scenario $|J_0| \lesssim K_0$, $\gamma_S(T_{\rm K}^K)$ is controlled by $I^*_+$ thus	 the spin susceptibility crosses over to strong coupling concomitantly with $\chi_T$. However, in the strongly ferromagnetic case $|J_0| \gg K_0$, the retardation of $I\to I^*_+$ over $K\to K^*$ leads to a crossover of $\chi_S$ at even lower temperatures. These findings are consistent with the numerical results of~\cite{KotliarHauleYin} and provide a simple picture of the incoherent regime of Hund's metals at intermediate energy scales: composite quasiparticles incorporate orbital degrees of freedom but not spin degrees of freedom, screening $\boldsymbol{T}$ but not $\boldsymbol{S}$. 

\paragraph{Discussion.}
We studied impurities in the presence of strong Hund's coupling in terms of a Kondo problem with spin and orbital degrees of freedom. The spin coupling can be ferromagnetic or antiferromagnetic depending on the filling of the underlying Anderson impurity model. In the Hund's metal region, very close to half-filling the coupling is ferromagnetic and this is the regime that corresponds to hole-doped iron pnictides, while the antiferromagnetic case is realized in the strongly electron-doped regime.
In the ferromagnetic case, there is a subtle interplay of spin and orbital degrees of freedom which leads to protracted flows until the Fermi liquid.
This explains the strong doping dependence of the coherence scale that has been observed, the electron-doped iron pnictides such as Fe$_{1-x}$Co$_x$Ba$_2$As$_2$~\cite{Rullier} having a much larger coherence temperature than hole-doped materials such as KFe$_2$As$_2$~\cite{Hardy}.

Finally, our Kondo impurity model describes true magnetic impurities with large Hund's coupling and embedded in metallic hosts. Thus the above mechanism also applies to dilute transition-metal magnetic alloys and successfully reproduces the overall trend of the experimentally measured coherence temperature as a function of filling~\cite{Daybell}.

We are grateful to N. Andrei, {J. von Delft}, K. Haule, M. Kharitonov, {K. Stadler} and A.M. Tsvelik for insightful discussions. 
This work has been supported by NSF grant Nos. DMR-0906943 and DMR-115181.

\begin{widetext}
%\documentclass[aps,prl,onecolumn,superscriptaddress]{revtex4}
%\usepackage{amsmath}
%\usepackage{graphicx}
%\usepackage{dcolumn}
%\usepackage{bm}
%\usepackage{amsfonts}
%\makeatletter
%\def\captionof#1#2{{\def\@captype{#1}#2}}
%\makeatother
%
%\newcommand{\rme}{{\rm e}}
%\newcommand{\rmd}{{\rm d}}
%\newcommand{\rmi}{{\rm i}}
%\newcommand{\ud}[1]{\hspace{-0em}\mathrm{d}{#1}\;}
%\newcommand{\udd}[2]{\hspace{-0em}\mathrm{d}{#1}\,\mathrm{d}{#2}\;}
%\newcommand{\Ud}[1]{\hspace{-0.5ex}\mathrm{d}{#1}\;}
%\newcommand{\Udd}[2]{\hspace{-0.5ex}\hspace{-0em}\mathrm{d}{#1}\,\mathrm{d}{#2}\;}
%\newcommand{\uD}{\mathcal{D}}
%
%\begin{document}

\section{Supplementary material}

\begin{center}
Camille Aron, Gabriel Kotliar
\end{center}

\bigskip

In this note, we give some details on the derivation of our SU($M$)$\times$SU($N$) Kondo model starting from an SU($M$)$\times$SU($N$) Anderson impurity model, both in the presence of a strong Hund's coupling as well as in the absence of Hund's coupling ($J_{\rm H} = 0$). We also detail some group-theoretic aspects of the derivation of the weak-coupling RG equations. Finally, we derive the scaling equations governing the temperature dependence of the spin and orbital static susceptibilities.

To simplify, we replace the notation for the impurity valence $n_d \mapsto n$ throughout this note.
\subsection{Schrieffer-Wolff transformation}
We recall the definition of the Anderson Hamiltonian introduced in Eqs.~(1)-(3) of the Letter:
\begin{align}
H_{\rm AIM} =& H_{\rm imp} + H_{\rm hyb} + H_{\rm bath} \;,
\end{align}
with 
\begin{align}
H_{\rm imp} &\equiv  \epsilon_d n + \frac{1}{2} U n(n-1)
 + \frac{1}{2} J_{\rm H} \sum_{mn,\, \sigma\sigma' }  d_{m\sigma}^\dagger d_{n\sigma'}^\dagger d_{m\sigma'} d_{n\sigma}\;, \\
H_{\rm hyb} &\equiv V  \sum_{k,m,\sigma}    \psi^\dagger_{km\sigma} d_\sigma + {\rm h.c.} \;,  \\
H_{\rm bath} &\equiv  \sum_{k,m,\sigma} \epsilon_k \psi^\dagger_{km\sigma} \psi_{km\sigma}\;. 
\end{align}
$\sigma,\sigma' = 1\ldots N$ are SU($N$) spin indices, $m,n = 1\ldots M$ are SU($M$) orbital indices, 
and $U \gg J_{\rm H} > 0$.
Notice that the Hund's coupling can be recast as
\begin{align}
 \frac{1}{2} J_{\rm H} \sum_{mn,\, \sigma\sigma' }  d_{m\sigma}^\dagger d_{n\sigma'}^\dagger d_{m\sigma'} d_{n\sigma} = J_{\rm H} \left[  \frac{nN}{2} - \frac{n^2}{2N} -  \left( \sum_m \boldsymbol{s}_m\right)^2 \right]\;,
 \end{align}
 where $n\equiv \sum_{m\sigma} d_{m\sigma}^\dagger d_{m\sigma}$ and $\boldsymbol{s}_m \equiv \sum_{\sigma\sigma'}  d_{m\sigma}^\dagger \frac{\boldsymbol{\sigma}_{\sigma\sigma'}}{2}d_{m\sigma'}$.
 % are SU($N$)-spins in their fundamental representation and  $\boldsymbol{s}_m^2 = \frac{N^2-1}{2N}$.
For a given valence $n$, the lowest energy level of  $H_{\rm imp}$ corresponds to the situation in which the $n$ SU($N$)-spins are aligned and
\begin{align}
 E_{n} = n \epsilon_d + \frac{1}{2} n(n-1) [U -  J_{\rm H} {(1/N +1/2)} ] \;.
\end{align}
The goal is to re-write $ H_{\rm AIM}$ as the Kondo model $H_\mathrm{K} = H_\mathrm{int} +  H_\mathrm{bath}$ with
\begin{align}
 H_{\rm int} =& J_{\rm p} \, \left(\mathbb{I}_T \otimes\mathbb{I}_S \right) \otimes \left( \psi_{m\sigma}^\dagger \psi_{m\sigma} \right)  \nonumber \\
& + J_0 \, \left( \mathbb{I}_T \otimes\boldsymbol{S} \right) \otimes \left( \psi_{m\sigma} \frac{\boldsymbol{\sigma}_{\sigma \sigma'}}{2} \psi_{m\sigma'} \right)  \nonumber \\
& + K_0 \,\left( \boldsymbol{T}   \otimes\mathbb{I}_S \right) \otimes \left( \psi_{m\sigma} \frac{\boldsymbol{\tau}_{mn}}{2}  \psi_{n\sigma} \right)  \nonumber \\
& + I_0 \, \left( \boldsymbol{T} \otimes\boldsymbol{S}   \right) \otimes
 \left( \psi_{m\sigma} \frac{\boldsymbol{\sigma}_{\sigma \sigma'}}{2} \frac{\boldsymbol{\tau}_{mn}}{2} \psi_{n\sigma'} \right)\;.
\end{align}
The Schrieffer-Wolff transformation (second order perturbation theory in the hybridization) is given by
\begin{align}
 H_{\rm int} \simeq - P_{n} H_{\rm hyb} \left[ \frac{P_{n+1}}{\Delta E^+}  + \frac{P_{n-1}}{\Delta E^-} \right] H_{\rm hyb} P_{n}\;,
\end{align}
where $P_{n\pm 1}$ are the projectors on the Hilbert space of valence $n\pm 1$ and 
\begin{align}
 \Delta E^+ &\equiv E_{n+1} - E_{n}   =  \epsilon_d + n [U - J_{\rm H}  {(1/N +1/2)} ] \simeq  \epsilon_d + n U\;, \\
 \Delta E^- &\equiv E_{n-1} - E_{n} = - \epsilon_d - (n-1) [U - J_{\rm H}  {(1/N +1/2)}] \simeq - \epsilon_d - (n-1) U\;,
\end{align}
where we used the fact that $U \gg J_{\rm H}$ to simplify the expressions.
Notice that $ \Delta E^+  +  \Delta E^- =  U > 0$. The conditions $\Delta E^+ > 0$ and $\Delta E^- > 0$ are simultaneously fulfilled for
\begin{align*}
\epsilon_d = -(n-1 + \alpha)  U  \mbox{ with } \alpha \in\, ]0,1[\;.
\end{align*}
$\alpha \to 1$ favors the virtual processes to the states with valence $n+1$ where as $\alpha \to 0$ favors the virtual processes to the states with valence $n-1$.

%\noindent [Check for simple half-filled Kondo model. $n=1/2$: $\epsilon_d \in [-U/2, U/2]$]

\medskip

\noindent Next, we use the following orthonormal basis of SU($M$)$\times$SU($N$)
\begin{align}
I^\alpha \equiv 
\left\{\begin{array}{l}
\frac{1}{\sqrt{M}} \mathbb{I}_T \otimes \frac{1}{\sqrt{N}} \mathbb{I}_S \\
\frac{1}{\sqrt{M}} \mathbb{I}_T \otimes \frac{1}{\sqrt{2}} \boldsymbol{\sigma} \\
\frac{1}{\sqrt{2}} \boldsymbol{\tau} \otimes \frac{1}{\sqrt{N}} \mathbb{I}_S \\
\frac{1}{\sqrt{2}} \boldsymbol{\tau} \otimes \frac{1}{\sqrt{2}} \boldsymbol{\sigma}
\end{array}
\right.
\end{align}
and the completeness relation
\begin{align}
\delta_{il}\delta_{jk} = \sum_\alpha I^{\alpha}_{ij} I^{\alpha}_{kl}\;,
\end{align}
where we gathered spin and orbital indices into $i \equiv (m,\sigma)$ and we sum over indices repeated twice, 
to re-write
\begin{align}
H_{\rm eff} =& 
-\left\{ \frac{V^2}{\Delta E^+} 
\left[ I^\alpha_{kl} P_n d_l P_{n+1} d_k^\dagger P_n \right]
\left[ \psi^\dagger_i I_{ij}^\alpha \psi_j \right] \right. \nonumber \\
& \left.  \qquad\qquad -\frac{V^2}{\Delta E^-} 
\left[ I^\alpha_{kl} P_n d_k^\dagger P_{n-1} d_l P_n \right]
\left[ \psi^\dagger_i I_{ij}^\alpha \psi_j \right]  \right\}\;.
\end{align}
The invariance of the Hamiltonian with respect to rotations of the spin and the orbital isospin implies that we can simply identify $J_{\rm p}$, $J_0$, $K_0$ and $I_0$ by computing single matrix elements:
\begin{align}
-\frac{1}{MN} 
  \left[
  \frac{V^2}{\Delta E^+}
  \langle \Phi | d_{m\sigma} P_{n+1} d^\dagger_{m\sigma} | \Phi \rangle
  -   \frac{V^2}{\Delta E^-}
    \langle \Phi | d^\dagger_{m\sigma} P_{n-1} d_{m\sigma} | \Phi \rangle
   \right]
   &= J_{\rm p} \langle \Phi| \mathbb{I}_T \otimes \mathbb{I}_S | \Phi \rangle \;, \label{eq:aJp}\\
-\frac{1}{M} {\sigma^{\alpha_0}_{\sigma\sigma'}}
  \left[
  \frac{V^2}{\Delta E^+}
  \langle \Phi | d_{m\sigma'} P_{n+1} d^\dagger_{m\sigma} | \Phi \rangle
  -   \frac{V^2}{\Delta E^-}
    \langle \Phi | d^\dagger_{m\sigma} P_{n-1} d_{m\sigma'} | \Phi \rangle
   \right]
   &= J_0 \langle \Phi| \mathbb{I}_T \otimes S^{\alpha_0} | \Phi \rangle \;, \label{eq:aJ0} \\
-\frac{1}{N} {\tau^{\alpha'_0}_{mn}} 
  \left[
  \frac{V^2}{\Delta E^+}
  \langle \Phi | d_{n\sigma} P_{n+1} d^\dagger_{m\sigma} | \Phi \rangle
  -   \frac{V^2}{\Delta E^-}
    \langle \Phi | d^\dagger_{m\sigma} P_{n-1} d_{n\sigma} | \Phi \rangle
   \right]
   &= K_0 \langle \Phi| T^{a_0} \otimes \mathbb{I}_S | \Phi \rangle \;, \label{eq:aK0} \\
- {\sigma^{\alpha_0}_{\sigma\sigma'}}  {\tau^{ a_0}_{mn}}
  \left[
  \frac{V^2}{\Delta E^+}
  \langle \Phi | d_{n\sigma'} P_{n+1} d^\dagger_{m\sigma} | \Phi \rangle
  -   \frac{V^2}{\Delta E^-}
    \langle \Phi | d^\dagger_{m\sigma} P_{n-1} d_{n\sigma'} | \Phi \rangle
   \right]
   &= I_0 \langle \Phi | T^{a_0} \otimes S^{\alpha_0} | \Phi \rangle \;, \label{eq:aI0}
\end{align}
where for each identification, we are free to choose a convenient state $|\Phi \rangle$ in the Hibert space of valence $n$, as well as the components $\alpha_0 \in \{ 1,\ldots, N^2-1 \}$ and $ a_0 \in \{ 1,\ldots, M^2-1\}$.

\subsection{Computation of the Kondo couplings}

\paragraph{\textbf{SU($N$) conventions.}} Henceforth, we denote the weights of the fundamental representation of SU($N$) from the highest to the lowest by $1, 2, \ldots, N$. The elements of the Cartan sub-algebra of SU($N$) are labelled correspondingly: $\sigma^1, \ldots, \sigma^{N-1}$, with $\mbox{Tr} [\sigma^{\alpha_0} \sigma^{\beta_0}] = 2\delta_{\alpha_0 \beta_0}$ where $\alpha_0, \beta_0 = 1, \ldots, N-1$.

\noindent Typically, we choose the state
\begin{align}
|\Phi \rangle = | \overbrace{\underbrace{1 \ldots 1}_{n} 0 \ldots 0 }^{M}\rangle\;.
\end{align} 
``$1$'' labels the highest weight of the fundamental representation of SU($N$) [$1=\uparrow$ along $\sigma^z$ in SU(2)] and ``$0$'' labels a vacant orbital. $|\Phi \rangle$ is normalized $\langle \Phi | \Phi \rangle=1 $, automatically anti-symmetrized in the orbital indices and it is symmetric by permutations of the spins since they are all ``$1$''.
We also choose $\alpha_0$ and $a_0$ so that $\sigma^{\alpha_0}$ and $\tau^{ a_0}$ are elements of the Cartan sub-algebras of SU($N$) and SU($M$). This simplifies greatly the analysis because those elements can be represented by $N\times N$ (or $M\times M$) diagonal matrices.

We now detail the computation of $J_0$, sketch the computation of $K_0$, and leave the computation of $J_{\rm p}$ and $I_0$ to the reader since they go along the same lines as for $J_0$.

\subsubsection{Computation of $J_0$}
To compute $J_0$, we need to compute both sides of
\begin{align}\label{eq:jeq}
-\frac{1}{M} {\sigma^{\alpha_0}_{\sigma\sigma'}}
  \left[
  \frac{V^2}{\Delta E^+}
  \langle \Phi | d_{m\sigma'} P_{n+1} d^\dagger_{m\sigma} | \Phi \rangle
  -   \frac{V^2}{\Delta E^-}
    \langle \Phi | d^\dagger_{m\sigma} P_{n-1} d_{m\sigma'} | \Phi \rangle
   \right]
   &= J_0 \langle \Phi| \mathbb{I}_T \otimes S^{\alpha_0} | \Phi \rangle \;.
\end{align}
We choose  $\alpha_0 = 1$ and work with the 'first' element of the Cartan sub-algebra of SU($N$),  which is the diagonal $N\times N$ matrix reading
\begin{align}
\frac{\sigma^{1}}{2} = \frac{1}{2}
\left( 
\begin{array}{ccccc}
1 &  & & &\\
 & -1 & & &\\
  &  & 0 & & \\
    &  &  &  \ddots & \\
    &  &  & & 0
\end{array}
\right)\;,
\end{align}
[for SU(2), it is $\frac{\sigma^z}{2}$]
and we choose the state
\begin{align}
| \Phi \rangle = | \overbrace{\underbrace{1 \ldots 1}_{n} 0 \ldots 0 }^{M}\rangle \;.
\end{align}
It is normalized $\langle \Phi | \Phi \rangle=1 $, automatically anti-symmetrized in the orbital indices and it is symmetric by permutations of the spins since they are all $1$.

\paragraph{\textbf{RHS of  Eq.~(\ref{eq:jeq}).}}
$| \Phi \rangle$ is the state with highest weight of our totally symmetric spin representation and the corresponding value of $S^1$ is simply
\begin{align}
\langle \Phi |  S^{1}| \Phi \rangle  = n \times \frac{1}{2}\;,
\end{align}
so that the {\sc rhs} of Eq.~(\ref{eq:jeq}) is
\begin{align}
J_0 \langle \Phi| \mathbb{I}_T \otimes S^{\alpha_0} | \Phi \rangle  = \frac{n}{2} J_0\;.
\end{align}

\paragraph{\textbf{LHS of  Eq.~(\ref{eq:jeq}).}}
Let us start with the term in $1/\Delta E^+$. The sum over $\sigma$ and $\sigma'$ is simplified because the matrix $\sigma^1$ is diagonal, and the only non-vanishing elements are $\sigma=\sigma'=1,2$. We have
\begin{align}
& -\frac{1}{M} {\sigma^{\alpha_0}_{\sigma\sigma'}}
  \frac{V^2}{\Delta E^+}
  \langle \Phi | d_{m\sigma'} P_{n+1} d^\dagger_{m\sigma} | \Phi \rangle \nonumber \\
&\qquad = -\frac{1}{M} 
  \frac{V^2}{\Delta E^+}
  \left[ 
  \langle \Phi | d_{m1} P_{n+1} d^\dagger_{m1} | \Phi \rangle
-    \langle \Phi | d_{m2} P_{n+1} d^\dagger_{m2} | \Phi \rangle
\right] \\
&\qquad = 
-\frac{1}{M} 
  \frac{V^2}{\Delta E^+}
  \left[ 
\sum_{m=n+1}^M
\langle  \overbrace{1 \ldots 1}^{n} 0 \ldots 0 \underbrace{1}_{\mathrm{pos=}m} 0 \ldots 0 
| P_{n+1} | 
\overbrace{1 \ldots 1}^{n} 0 \ldots 0 \underbrace{1}_{\mathrm{pos}=m} 0 \ldots 0   \rangle \right. \nonumber \\
& \qquad\qquad 
\left.
-
\sum_{m=n+1}^M
\langle  \overbrace{1 \ldots 1}^{n} 0 \ldots 0 \underbrace{2}_{\mathrm{pos}=m} 0 \ldots 0 
| P_{n+1} | 
\overbrace{1 \ldots 1}^{n} 0 \ldots 0 \underbrace{2}_{\mathrm{pos}=m} 0 \ldots 0   \rangle 
\right] \;. \label{eq:prepJ}
\end{align}
The relevant states contributing to the projector on the Hilbert space of valence $n+1$, $P_{n+1}$, are
\begin{align}
& | \Psi_{m} \rangle \equiv | \overbrace{1 \ldots 1}^{n} 0 \ldots 0 \underbrace{1}_{\mathrm{pos}=m} 0 \ldots 0   \rangle, \qquad  m=n+1\ldots M\;,\\
&| \Psi'_{m} \rangle \equiv \frac{1}{\sqrt{n+1}} \sum_{i\in \{1\ldots n,m \}} 
| \overbrace{ 1 \ldots 1 \underbrace{2}_{\mathrm{pos}=i} 1 \ldots 1}^{n} 0 \ldots 0 \underbrace{1}_{\mathrm{pos}=m} 0 \ldots 0   \rangle, \qquad  m=n+1\ldots M\;.
\end{align}
These states are totally antisymmetric in the orbital sector and totally symmetric in the spin sector and we were careful to normalize them properly. Performing the projection in Eq.~(\ref{eq:prepJ}) with
\begin{align}
P_{n+1} = \sum_{m=n+1}^{M} | \Psi_{m} \rangle \langle \Psi_{m} | + | \Psi'_{m} \rangle \langle \Psi'_{m} |  + \ldots
\end{align}
we obtain
\begin{align}
 -\frac{1}{M} {\sigma^{\alpha_0}_{\sigma\sigma'}}
  \frac{V^2}{\Delta E^+}
  \langle \Phi | d_{m\sigma'} P_{n+1} d^\dagger_{m\sigma} | \Phi \rangle
=
-\frac{1}{M} 
  \frac{V^2}{\Delta E^+}
  \left[ 
(M-n) - (M-n)  \frac{1}{n+1}
\right] \;.
\end{align}

Let us now cope with the term in $1/\Delta E^-$ in Eq.~(\ref{eq:jeq}). Here also, the sum over $\sigma$ and $\sigma'$ is simplified because the matrix $\sigma^1$ is diagonal, and only the element $\sigma=\sigma'=1$ contribute to the sum (the element $\sigma=\sigma'=2$ is irrelevant since there is no spin $2$ to annihilate in $|\Phi\rangle$).
 We have
\begin{align}
& \frac{1}{M} {\sigma^{\alpha_0}_{\sigma\sigma'}}
     \frac{V^2}{\Delta E^-}
    \langle \Phi | d^\dagger_{m\sigma} P_{n-1} d_{m\sigma'} | \Phi \rangle
\nonumber \\
&\quad =\frac{1}{M} 
  \frac{V^2}{\Delta E^-}
  \langle \Phi | d_{m1} P_{n-1} d^\dagger_{m1} | \Phi \rangle \\
&\quad = 
\frac{1}{M} 
  \frac{V^2}{\Delta E^-}
\sum_{m=1}^{n}
\langle  \overbrace{1 \ldots 1 \underbrace{0}_{\mathrm{pos}=m} 1 \ldots 1}^{n} 0 \ldots 0 | P_{n-1} | 
\overbrace{1 \ldots 1 \underbrace{0}_{\mathrm{pos}=m} 1 \ldots 1}^{n} 0 \ldots 0  \;. \rangle\label{eq:prepJ2}
\end{align}
The relevant states contributing to the projector to the Hilbert space of valence $n-1$, $P_{n-1}$, are
\begin{align}
| \Psi_m \rangle \equiv |\overbrace{1 \ldots 1 \underbrace{0}_{\mathrm{pos}=m} 1 \ldots 1}^{n} 0 \ldots 0  \rangle,\qquad  m=1\ldots n\;.
\end{align}
Performing the projection in Eq.~(\ref{eq:prepJ2}) with
\begin{align}
P_{n-1} = \sum_{m=1}^{n} | \Psi_m \rangle \langle \Psi_m | + \ldots
\end{align}
we obtain
\begin{align}
\frac{1}{M} {\sigma^{\alpha_0}_{\sigma\sigma'}}
     \frac{V^2}{\Delta E^-}
    \langle \Phi | d^\dagger_{m\sigma} P_{n-1} d_{m\sigma'} | \Phi \rangle = \frac{1}{M} 
  \frac{V^2}{\Delta E^-}
n  \;.
\end{align}
Altogether, summing the terms in $1/\Delta E^+$ and $1/\Delta E^-$, we obtain
\begin{align}
-\frac{1}{M} 
\left[ 
\frac{M-n}{n+1}
  \frac{n}{\Delta E^+}
- n 
 \frac{1}{\Delta E^-}
\right] V^2 = \frac{n}{2} J_0 \;,
\end{align}
and finally,
\begin{align}
J_0 = \frac{1}{2M} \left[ 
 \frac{1}{\Delta E^-} - \frac{M-n}{n+1}
  \frac{1}{\Delta E^+} \right] V^2\;,
\end{align}

\subsubsection{Computation of $K_0$}
To compute $K_0$, we need to compute both sides of
\begin{align} \label{eq:keq}
-\frac{1}{N} {\tau^{a_0}_{mn}} 
  \left[
  \frac{V^2}{\Delta E^+}
  \langle \Phi | d_{n\sigma} P_{n+1} d^\dagger_{m\sigma} | \Phi \rangle
  -   \frac{V^2}{\Delta E^-}
    \langle \Phi | d^\dagger_{m\sigma} P_{n-1} d_{n\sigma} | \Phi \rangle
   \right]
   &= K_0 \langle \Phi| T^{a_0} \otimes \mathbb{I}_S | \Phi \rangle \;,
\end{align}
We choose  $a_0 = n$ and work with the $n^\mathrm{th}$ element of the Cartan sub-algebra which is the diagonal $N\times N$ matrix reading
\begin{align}
\frac{\sigma^{n}}{2} = \frac{1}{\sqrt{2n(n+1)}}
\left( 
\begin{array}{ccccccc}
1 &  & & & & &\\
 & \ddots & & & & &\\
 & & 1 & & & &\\
  &  & &  -n & & &\\
   &  & &  &  0 & &\\
    &  &  &  &  & \ddots & \\
     &  &  &  &  & & 0
\end{array}
\right)\;,
\end{align}
in which the element of value $-n$ is located at the $n+1^\mathrm{th}$ position.
and the state
\begin{align}
| \Phi \rangle = | \overbrace{\underbrace{1 \ldots 1}_{n} 0 \ldots 0 }^{M}\rangle \;.
\end{align}
It is normalized $\langle \Phi | \Phi \rangle=1 $, automatically anti-symmetrized in the orbital indices and it is symmetric by permutations of the spins since they are all $1$.

\paragraph{\textbf{RHS of  Eq.~(\ref{eq:keq}).}}
The weight of the state $|\Phi\rangle$ on the direction $T^n$ can be computed as
\begin{align}
\langle \Phi |  T^{n}| \Phi \rangle  = \frac{n}{\sqrt{2n(n+1)}}\;,
\end{align}
so that the {\sc rhs} of Eq.~(\ref{eq:keq}) is
\begin{align}
K_0 \langle \Phi| T^{a_0} \otimes \mathbb{I}_S | \Phi \rangle  = \frac{n}{\sqrt{2n(n+1)}} K_0\;.
\end{align}

\paragraph{\textbf{LHS of  Eq.~(\ref{eq:keq}).}} We leave this exercise to the reader since the computation goes along the same lines as for $J_0$.

\medskip

Repeating the exercise for $J_{\rm p}$ and $I_0$, we identify
\begin{align}
 J_{\rm p} &= \frac{1}{MN} \left[ \frac{n}{\Delta E^-} - \frac{M-n}{n+1} \frac{N+n}{\Delta E^+}    \right] V^2\;, \\
 J_0 &= \frac{2}{M} \left[ \frac{1}{\Delta E^-} - \frac{M-n}{n+1} \frac{1}{\Delta E^+}  \right] V^2 \;, \\
 K_0 &= \frac{2}{N} \left[ \frac{1}{\Delta E^-} + \frac{N+n}{n+1} \frac{1}{\Delta E^+}   \right] V^2 \;, \\
 I_0 &= 4 \left[ \frac{1}{n} \frac{1}{\Delta E^-}+   \frac{1}{n+1} \frac{1}{\Delta E^+}\right] V^2  \;.
\end{align}

\subsection{Case without Hund's coupling, $J_{\rm H} = 0$}
In the absence of Hund's coupling, since all states at a given valence are degenerate, the projectors $P_{n+1}$ and $P_{n-1}$ in the expressions of Eqs.~(\ref{eq:aJp})-(\ref{eq:aI0}) play no role and this simplifies greatly the analysis. Contrary to the case of a strong Hund's coupling, all physical configurations are now allowed.
We briefly detail the case of $J_0$ and leave the computation of the other couplings as an exercise.

Without the projectors $P_{n\pm 1}$, Eq.~(\ref{eq:aJ0}) reduces to
\begin{align}
-\frac{1}{M} {\sigma^{\alpha_0}_{\sigma\sigma'}}
  \left[
  \frac{V^2}{\Delta E^+}
  \langle \Phi | d_{m\sigma'} d^\dagger_{m\sigma} | \Phi \rangle
  -   \frac{V^2}{\Delta E^-}
    \langle \Phi | d^\dagger_{m\sigma} d_{m\sigma'} | \Phi \rangle
   \right]
   &= J_0 \langle \Phi| \mathbb{I}_T \otimes S^{\alpha_0} | \Phi \rangle \;.
\end{align}
Similarly to what we did in the Hund's case above, we choose $\alpha_0 = 1$ (\textit{i.e.} working with $\sigma^1$, the first element of the Cartan sub-algebra) and the state $|\Phi \rangle = | \overbrace{\underbrace{1 \ldots 1}_{n} 0 \ldots 0 }^{M}\rangle$. Using $\sum_{m=1}^M \langle \Phi |  d_{m1} d^\dagger_{m1}   |\Phi \rangle  = M - n$, $\sum_{m=2}^M \langle \Phi |  d_{m1} d^\dagger_{m2}   |\Phi \rangle  = M$, $\sum_{m=1}^M \langle \Phi |  d^\dagger_{m1} d_{m1}   |\Phi \rangle  = n$, and $\sum_{m=1}^M \langle \Phi |  d^\dagger_{m2} d_{m2}   |\Phi \rangle  = 0$, we get
\begin{align}
-\frac{1}{M} 
  \left[-n
  \frac{1}{\Delta E^+}
  -   n\frac{1}{\Delta E^-}
   \right] V^2
   &= \frac{J_0}{2} n\;,
\end{align}
and therefore
\begin{align}
J_0 &= \frac{2}{M} \left[ \frac{1}{\Delta E^+} + \frac{1}{\Delta E^-} \right] V^2\;.
\end{align}
Similarly, we find
\begin{align}
J_{\rm p} &= -\frac{1}{MN} \left[ \frac{MN-n}{\Delta E^+} - \frac{n}{\Delta E^-} \right] V^2\;, \\
K_0 &= \frac{2}{N} \left[ \frac{1}{\Delta E^+} + \frac{1}{\Delta E^-} \right] V^2\;, \\
I_0 &= \frac{4}{n} \left[ \frac{1}{\Delta E^+} + \frac{1}{\Delta E^-} \right] V^2\;,
\end{align}
\textit{i.e.} $n_d I_0 = 2 M J_0 = 2 N K_0$. The model reduces to an SU($M\times N$) Coqblin-Schrieffer impurity model, with a single Kondo coupling, in which the spin lives in the totally anti-symmetric representation of SU($M\times N$), composed of $n_d$ impurity electrons (single column Young tableau).

\subsection{RG: 	Projection of terms $I^2  \left(\boldsymbol{S} \cdot \boldsymbol{\sigma} \right) \left( \boldsymbol{Q}\cdot \boldsymbol{\tau} \right)$ on $\left( \boldsymbol{S}  \cdot \boldsymbol{\sigma} \right) \left( \boldsymbol{T} \cdot \boldsymbol{\tau} \right) $}
In this section, we provide the computational details needed to perform the projection of terms such as  $I^2  \left(\boldsymbol{S} \cdot \boldsymbol{\sigma} \right) \left( \boldsymbol{Q}\cdot \boldsymbol{\tau} \right)$, with $Q^c \equiv \{T^a,T^b \} \mbox{Tr}\left[ \tau^{\{a} \tau^b \tau^{c\}} \right]$,
 on the original model Hamiltonian, namely on a term  $ b \left( \boldsymbol{S}  \cdot \boldsymbol{\sigma} \right) \left( \boldsymbol{T} \cdot \boldsymbol{\tau} \right)$ where $b$ is a constant to determine which depends \textit{a priori} on the representation of the orbital isospin $\boldsymbol{T}$.
 
\paragraph{\textbf{SU($N$) conventions.}}
$S^\alpha$ ($\alpha=1 \ldots N^2-1$) and $T^a$ ($a= 1 \ldots M^2-1$) are the respective generators of the SU($N$) and SU($M$) group in their respective generic representations $\{S\}$ and $\{T\}$.
They obey the Lie algebra defining commutation relations
\begin{align}
 [S^\alpha;S^\beta] = \rmi f_{\alpha\beta\gamma} S^\gamma\;, \qquad
 [T^a;T^b] = \rmi f_{abc} T^c \;.
\end{align}
$\sigma^\alpha$ and $\tau^a$ are the respective generators of  SU($N$) and SU($M$) in their fundamental representations that we denote $\{\sigma\}$ and $\{\tau\}$. They also obey the Lie algebra defining commutation relations
\begin{align}
 \left[\frac{\sigma^\alpha}{2};\frac{\sigma^\beta}{2}\right] = \rmi f_{\alpha\beta\gamma} \frac{\sigma^\gamma}{2}\;, \qquad
 \left[\frac{\tau^a}{2};\frac{\tau^b}{2}\right] = \rmi f_{abc} \frac{\tau^c}{2} \;,
\end{align}
where $f_{abc}$  and $f_{\alpha\beta\gamma}$ are  completely antisymmetric tensors. We choose them to be normalized  such that 
\begin{align}
\mbox{Tr}\left[\sigma^\alpha \sigma^\beta \right] = 2 \delta_{\alpha \beta} \mbox{ and } \mbox{Tr}\left[\tau^a \tau^b \right] = 2 \delta_{ab}\;.
\end{align}
Furthermore, they have the property (special to the fundamental representations)
\begin{align}
 \left\{\frac{\sigma^\alpha}{2};\frac{\sigma^\beta}{2}\right\} =  d_{\alpha\beta\gamma} \frac{\sigma^\gamma}{2} + \frac{1}{N} \delta_{\alpha\beta} \mathbb{I}_\sigma \;, \qquad
  \left\{ \frac{\tau^a}{2};\frac{\tau^b}{2} \right\}  = d_{abc} \frac{\tau^c}{2} + \frac{1}{M} \delta_{ab} \mathbb{I}_\tau\;,
\end{align}
where $d_{abc}$ and $ d_{\alpha\beta\gamma}$ are totally symmetric tensors (so-called the structure tensors of the Lie algebras). These two sets of properties lead to the following multiplication laws
\begin{align}
 \frac{\sigma^a}{2} \frac{\sigma^b}{2} &= \frac{\mathbb{I}_\sigma}{2N} \delta_{ab} + \frac{1}{2} \left( d_{abc} + \rmi f_{abc} \right) \frac{\sigma^c}{2} \\
 \frac{\tau^\alpha}{2} \frac{\tau^\beta}{2} &= \frac{\mathbb{I}_\tau}{2M} \delta_{\alpha\beta} +  \frac{1}{2} \left( d_{\alpha\beta\gamma} + \rmi f_{\alpha\beta\gamma} \right)\ \frac{\tau^\gamma}{2} \,
\end{align}
that used iteratively can decomposed any product of generators $\sigma^{\alpha_1} \sigma^{\alpha_2} \ldots \sigma^{\alpha_n}$ onto the complete basis formed by $\mathbb{I}_\sigma$ and the $\sigma^\alpha$'s. 

The Casimir invariants are defined as 
\begin{align}
C_n^S  \, \mathbb{I}_S \equiv \mbox{Tr}[ \frac{\sigma^{\{ \alpha_1}}{2} \ldots  \frac{\sigma^{\alpha_n \}}}{2}] S^{\alpha_1} \ldots S^{\alpha_n}\;,
\end{align}
where $\mathbb{I}_S$ is the identity in the representation of $\boldsymbol{S}$ and $\{a_1 \ldots a_n \}$ stands for the sum over all permutations weighted by $1/n!\,$. 
 
\paragraph{\textbf{Identification of $b$.}}
Let us  work with the matrix $Q^c \equiv  \{T^a;T^b \} d_{abc}$. 
It is an Hermitian matrix, therefore $\mbox{Tr}\left[ (Q^c)^2 \right] > 0$. A sub-basis of the vector space of such Hermitian matrices is given by $\left\{ \mathbb{I}_{T}, T^a (a = 1 \ldots M^2 -1) \right\}$. 
The following scalar product can be given to that  sub-vector space:
 $\langle Q^c | Q^\delta \rangle \equiv  \mbox{Tr}\left[ Q^c Q^\delta\right]$. Indeed one has
$\mbox{Tr}\left[ T^a T^b\right] \propto \delta_{ab}$. The basis can be completed to the full vector space by a set of traceless matrices  $U^k$  with $\mbox{Tr}\left[T^a  U^k \right] = 0$, $\mbox{Tr}\left[ U^k U^{k'}\right] \propto \delta_{kk'}$.
In full generality, $Q^c$ can be decomposed in this complete basis as
\begin{align} \label{eq:decompM}
 Q^c = a^c \mathbb{I}_\tau + b^{c}_a T^a + c^c_{k} U^k\;.
\end{align}
Now, in order for the expression $Q^e \tau^e$ to be a scalar (\textit{i.e.} a quantity that is invariant under the simultaneous rotations of the isospin of the impurity and the one of the conduction electrons) we must have the coefficients $a^c = 0$. Below, we check that indeed $a^c = 0$  and we identify $b^{c}_a$.
In order  to extract the coefficient $a^c$, we trace the above expression:
\begin{align}
\mbox{Tr}\left[  \{T^a;T^b \}  \right] d_{abc} &= a^{c} d_{\{T\}}\;,
\end{align}
where $ d_{\{T\}} $ is the dimension of the representation $\{T\}$. Using the relation
\begin{align} \label{eq:tr2}
  \mbox{Tr}\left[ T^a T^b \right] = \frac{1}{2} \frac{ d_{\{T\}} C_2(\{T\}) }{d_{\{\tau\}} C_2(\{\tau\}) }  \delta_{ab}\;,
\end{align}
where $ d_{\{\tau\}}=M $ is the dimension of the fundamental representation of SU($M$), and $C_2(\{T\})$ is the eigen-value of the second order Casimir in the representation $\{T\}$ [$C_2(\{T\})=\frac{M^2-1}{2M}$ is the eigen-value of the second order Casimir in the fundamental representation], we obtain
\begin{align}
\frac{ d_{\{T\}} C_2(\{T\}) }{d_{\{\tau\}} C_2(\{\tau\}) }  d_{aac}  &= a^{c}  d_{\{T\}}\;.
\end{align}
The {\sc lhs} is zero because the tensor $ d_{abc} $ is traceless, therefore we get $a^c = 0$.

In order to extract the coefficient $b^c$, we multiply Eq.~({\ref{eq:decompM}}) by $T^e$ and trace:
\begin{align} \label{eq:uneqaupif}
\mbox{Tr}\left[ Q^c T^e \right] = b^{c}_a \mbox{Tr}\left[ T^a T^e \right]\;.
\end{align}
Using the relation (\ref{eq:tr2}) we obtain
\begin{align}
  \mbox{Tr}\left[  \{T^a;T^b \}  T^e \right] d_{abc} = \frac{1}{2}
  b^{c}_e
  \frac{ d_{\{T\}} C_2(\{T\}) }{d_{\{\tau\}} C_2(\{\tau\}) }  \;.
\end{align}
One can also compute the {\sc{lhs}} of Eq.~(\ref{eq:uneqaupif}) as
\begin{align}
  \mbox{Tr}\left[  \{T^a;T^b \}  T^e \right] d_{abc} &= 2 \, \mbox{sTr} \left[ T^a,T^b,T^e \right]  d_{abc} 
\equiv 2 d^{\{T\}}_{abe} d_{abc}\;.
\end{align}
with the symmetrized trace defined as
\begin{align}
 \mbox{sTr} \left[ T^{a_1}, \ldots, T^{a_n} \right] \equiv \frac{1}{n!} \sum_\pi \mbox{Tr}\left[ T^{a_{\pi(1)}} \ldots T^{a_{\pi(n)}} \right]	\;.
\end{align}
where the sum is performed over all permutations of the indices. Using the following property (\textit{c.f.} Eq.~(114) in \cite{Ritbergen})
\begin{align}
 d^{\{T\}}_{abe} d_{abc} = \frac{1}{d_{\{\mathrm{Adj}\}}} d^{\{T\}}_{ab\delta} d_{ab\delta} \delta_{ce}\,
\end{align}
 where $d_{\{\mathrm{Adj}\}}=M^2-1$ is the dimension of the adjoint representation of SU($M$), together with the contraction identity (\textit{c.f.} Eq.~(A21) in \cite{Okubo})
\begin{align}
 d^{\{T\}}_{ab\delta} d_{ab\delta}  = d_{\{T\}} C_3(\{T\})\;,
\end{align}
[where $C_3(\{T\})$ is the eigen value of the third order Casimir -- in the fundamental representation, $C_3(\{\tau\}) = (M^2-1)(M^2-4)/4M^2$] we get
\begin{align}
 2  \frac{d_{\{T\}}}{d_{\{\mathrm{Adj}\}}} C_3(\{T\}) \delta_{ce} = \frac{1}{2} 
  b^{c}_e
  \frac{ d_{\{T\}} C_2(\{T\}) }{d_{\{\tau\}} C_2(\{\tau\}) }  \;,
\end{align}
implying that $b^{c}_e = b \delta_{ce} $ where $b$ depends on the representation and is given by
\begin{align}
 b(\{T\}) = 4  \frac{d_{\{\tau\}}}{d_{\{\mathrm{Adj}\}}}  \frac{ C_2(\{\tau\})}{ C_2(\{T\})}  C_3(\{T\}) = 2 \frac{ C_3(\{T\})}{ C_2(\{T\})}\;.
\end{align}
Notice that in the fundamental representation, $ C_3(\{\tau\}) = \frac{(M^2-1)(M^2-4)}{4M^2}$ so that $b(\{\tau\}) = (M^2-4)/M$.

{
\subsection{Scaling of static susceptibilities}
The finite-temperature impurity spin and orbital static susceptibilities are given respectively by
\begin{align}
\chi_S(T) = \int_0^{1/T} \ud{\tau} \langle \mathtt{T}\,  S^\alpha(\tau)  S^\alpha(0) \rangle \mbox{  and  }
\chi_T(T) = \int_0^{1/T} \ud{\tau} \langle \mathtt{T} \, T^a(\tau)  T^a(0) \rangle\;,
\end{align}
where $\tau$ is a Matsubara time and $\mathtt{T}$ is the Matsubara time-ordering operator.
To second order in the Kondo couplings we find
\begin{align}
\chi_S(T) \propto& \frac{1}{T} \left[1 -  \frac{MN}{2} \left( J^2  + \frac{C_2(T)}{2} I^2 \right)\ln(D/T)  - \mbox{non log terms in } J_i^2 \right]\;, \label{eq:chi1}\\
\chi_S(T) \propto& \frac{1}{T} \left[1 -  \frac{MN}{2} \left( K^2  + \frac{C_2(S)}{2} I^2 \right)\ln(D/T)  - \mbox{non log terms in } J_i^2 \right]\;. \label{eq:chi2}
\end{align}
The multiplicative RG equations for $\chi_S$ and $\chi_T$ read
\begin{align} \label{eq:chiTr}
\chi_{S/T}(T;J_i(D),D) =& Z_{S/T}(D,D')\,  \chi_{S/T}(T;J_i(D'),D') \;,
%\chi_T(T;J(D),D) =& Z_T(D,D') \chi_T(T;J(D'),D')
\end{align}
in which $D$ and $D'$ are two different ultraviolet cut-off and the $Z$'s are the renormalization factors which, up to the second order in the couplings, can be deduced from Eqs.~(\ref{eq:chi1}) and (\ref{eq:chiTr}) and read
\begin{align}
Z_S(D,D') =& 1 + \frac{MN}{2} \left( J^2  + \frac{C_2(T)}{2} I^2 \right) \ln(D/D')\;, \\
Z_T(D,D') =& 1 + \frac{MN}{2} \left( K^2  + \frac{C_2(S)}{2} I^2 \right) \ln(D/D')\;. 
\end{align}
Applying the operator $(D' / Z ) \ \rmd/\rmd D'$ to the Eqs.~(\ref{eq:chiTr}), they can be reshaped as
\begin{align}\label{eq:chi3}
D \frac{\partial \chi_{S/T}}{\partial D} + \sum_i \beta_i \frac{\partial \chi_{S/T}}{\partial J_i} + \gamma_{S/T} \chi_{S/T} &= 0\;,
%D \frac{\partial \chi_T}{\partial D} + \beta_i(J) \frac{\partial \chi_T}{\partial J_i} + \gamma_T(J) \chi_T &= 0
\end{align}
with the functions
\begin{align}
\gamma_S(J) \equiv \frac{\partial \ln Z_S}{\partial \ln D} = \frac{MN}{2} \left( J^2  + \frac{C_2(T)}{2} I^2 \right) 
\mbox{ and 	}
\gamma_T(J) \equiv \frac{\partial \ln Z_T}{\partial \ln D} = \frac{MN}{2} \left( K^2  + \frac{C_2(S)}{2} I^2 \right)\;.
\end{align}
The Eqs.~(\ref{eq:chi3}) are linear first-order partial differential equations that can be solved by the methods of characteristics. One can check that the cutoff variation
\begin{align}
\frac{\rmd }{\rmd \rho} \left( \exp\left[ \int_1^\rho \frac{\rmd x}{x} \gamma_{S/T}(J_i(x D_0)) \right] \chi_{S/T}(T;J_i(\rho D_0),\rho D_0 ) \right) = 0
\end{align}
provided that the renormalized couplings obey the characteristic equations set by the $\beta$ functions, $\rho \frac{\rmd J_i}{\rmd \rho} = \beta_i(J(\rho D_0))$ with the initial conditions at $\rho=1$ given by $J_i = J_0, K_0, I_0$. Therefore, the static susceptibilities at different cutoff scales $D$ and $D'$ are related by
\begin{align}
\chi_{S/T}(T;J_i(D),D) = \exp \left[ - \int_{D'}^{D}  \frac{\rmd \Delta}{\Delta} \gamma_{S/T}(J_i(\Delta))
 \right]
\chi_{S/T}(T;J_i(D'),D')\;.
\end{align}
Choosing $D'=T$ such as to get rid of the logarithmic terms in Eq.~(\ref{eq:chi1}), the following scaling equations for the static susceptibilities are obtained
\begin{align}
\chi_{S/T}(T;J(D_0),D_0) &\sim  \frac{1}{T} 
\exp \left[ - \int_{T}^{D_0}  \frac{\rmd D}{D} \gamma_{S/T}(J_i(D))
 \right] \;.
% \\
%\chi_T(T;J(D),D) & \sim 
%\exp \left[ \int_{D}^{T}  \frac{\rmd D}{D} \gamma_T(J(D))
% \right] \frac{1}{T} \;.
\end{align}
}

%\end{document}

\end{widetext}

\end{document}